\documentclass[aps,pre,preprint,groupedaddress,showpacs]{revtex4-1}
\usepackage[dvips]{graphicx}
\usepackage{textcomp}
\usepackage{amssymb}

\newcommand{\mc}{\multicolumn}
\begin{document}

\title{
\Large\bf The interface tension in the improved Blume-Capel model}

\author{Martin Hasenbusch}
\email[]{Martin.Hasenbusch@physik.hu-berlin.de}
\affiliation{
Institut f\"ur Physik, Humboldt-Universit\"at zu Berlin,
Newtonstr. 15, 12489 Berlin, Germany}

\date{\today}

\begin{abstract}
 We study interfaces with periodic boundary conditions in the low temperature
 phase of the improved Blume-Capel model on the simple cubic lattice. 
 The interface free energy is defined by the difference of the free energy
 of a system with anti-periodic boundary conditions in one of the directions
 and that of a system with periodic boundary conditions in all directions. 
 It is obtained by integration
 of differences of the corresponding internal energies over the inverse 
 temperature. 
 These differences can be computed efficiently by using a variance reduced
 estimator that is based on the exchange cluster algorithm. 
 The interface tension
 is obtained from the interface free energy  by using  predictions
 based on effective interface models.
 By using our numerical results for the interface tension $\sigma$ and the 
 correlation
 length  $\xi$ obtained in previous work, we determine the universal amplitude
 ratios $R_{2nd,+} = \sigma_0 f_{2nd,+}^2= 0.3863(6)$, 
$R_{2nd,-} = \sigma_0 f_{2nd,-}^2= 0.1028(1) $ and 
$R_{exp,-}=\sigma_0 f_{exp,-}^2= 0.1077(3)$. Our results
 are consistent with those obtained previously for the three-dimensional 
 Ising model, confirming the universality hypothesis.
\end{abstract}

\maketitle

\section{Introduction}
Interfaces appear in a large number of systems in soft
condensed matter physics, in chemistry and in biology.  
These interfaces separate for example the components of
a binary liquid mixture, or a liquid and its vapour. The  
behaviour of interfaces might be described by effective 
models such as the capillary wave model \cite{Cap}.  Via 
duality interfaces are related with strings in gauge theories.
In the last few years there has been fundamental progress 
in understanding the wide predictive power of effective models 
of strings. See for example \cite{BrMe16} and references therein.
A key feature in this discussion is the Lorentz invariance
of the gauge model, or in the case of the interface, 
the Galilean invariance of the underlying  three-dimensional
system. In the case of a spin model on a lattice, 
Galilean invariance is restored as the critical point is 
approached. In the present study we therefore focus on the 
neighbourhood of the critical point.
Note however that for  binary mixtures of fluids and
off-lattice models of such systems, Galilean invariance is  not
limited to criticality.

If the phase transition of a binary system is continuous,
it belongs to the universality class of the three-dimensional
Ising model.
In the neighbourhood of a continuous phase transition the behaviour of
various quantities is given by power laws. For example the correlation
length behaves as
\begin{equation}
\label{xic}
\xi = f_{\pm} |t|^{-\nu} \;(1 + a_{\pm} |t|^{\theta} + b t + ...) \;,
\end{equation}
where $t=\beta_c -\beta$ is the reduced temperature, $\nu$ the critical 
exponent of the correlation length, and $f_{\pm}$ the amplitude in
the low and high temperature phase.
Note that mostly $t= (T-T_c)/T_c$ is used as definition of the reduced 
temperature. The present choice is more convenient for our purpose.
Such power laws are affected by corrections. The leading confluent 
one comes with the exponent $\theta= \nu \omega \approx 0.5$
and the leading analytic correction is given by $b t$.
For reviews on critical phenomena see for example
\cite{WiKo,Fisher74,Fisher98,PeVi}.

Very recently the critical exponents of the three-dimensional Ising 
universality class have been computed very accurately by using the 
conformal bootstrap method \cite{Kosetal16}. In particular 
$\nu=1/y_t=0.6299709(40)$, 
obtained from $3 - y_t = \Delta_{\epsilon} = 1.412625(10)$.
And in table 2 of \cite{SiDu16} one finds
$\omega= \Delta_{\epsilon'}-3 = 0.82968(23)$  for the exponent of the 
leading correction. These estimates are consistent with, but more precise
than $\nu=0.63002(10)$ and $\omega=0.832(6)$ obtained from a Monte Carlo
study of the improved Blume-Capel model on the simple cubic lattice 
\cite{MHcritical}. 

 The interface tension $\sigma$ is the 
free energy per area of an interface in the thermodynamic limit. The 
interface free energy is, roughly speaking, the difference of the free
energies of a system with an interface and a corresponding system without an 
interface. For a precise definition see section \ref{defineFS} below.
In the neighbourhood of the critical point, the interface tension behaves as
\begin{equation}
\label{sigmac}
 \sigma = \sigma_0 (-t)^{\mu}  \;[1 + a_{\sigma} (-t)^{\theta} + \bar b t + ...] \;,
\end{equation}
where $\mu= 2\nu$. Dimensionless combinations of amplitudes are, following 
Renormalization Group (RG)-theory, universal. Here we shall study
\begin{equation}
R_{\pm} = \sigma_0 f_{\pm}^2 \;.
\end{equation}
Both the amplitudes of the exponential and the second moment correlation
length have been considered in the literature.  $R_{\pm}$ has been determined
for various experimental systems and has been computed by 
 using for example field theoretic methods. 
Accurate estimates have been obtained 
by using Monte Carlo simulations of the Ising model.

Studying the  improved Blume-Capel model, $a_{\pm} \approx 0$, 
eq.~(\ref{xic}), and 
$a_{\sigma} \approx 0$, eq.~(\ref{sigmac}), should simplify
the analysis of the data obtained for the correlation length and the 
interface tension.

Recently we demonstrated that the exchange cluster algorithm 
 \cite{ReMaCh98,ChMaRe98} can be employed to define 
variance reduced estimators of differences of observables measured in 
two slightly different systems \cite{mysphere,myfilm}. Here we apply this
idea to the interface energy.
In \cite{mytwopoint} we employed the exchange cluster algorithm to define 
a variance reduced estimator of the two-point function for systems with
a spontaneously broken $\mathbb{Z}_2$-symmetry. 
The numerical estimates obtained in 
\cite{mytwopoint} for the correlation length are used here to calculate 
$R_{\pm}$. 

The outline of the paper is the following. In the next section we shall define
the Blume-Capel model. We discuss the geometry of the systems that we simulate
and define the interface free energy.
Then we recall the exchange cluster algorithm and define the variance reduced
estimator of the difference in the 
internal energy between the anti-periodic and the periodic system. Next we
present our numerical results. We study the performance of the variance 
reduced estimator. We compute the interface tension for a large range
of inverse temperatures. Finally we determine estimates for the universal
amplitude combinations $R_{\pm}$. 

\section{The model}
\label{themodel}
As in previous work, we study the Blume-Capel model on the simple cubic lattice.
The bulk system, for a vanishing external field, is defined by the reduced 
Hamiltonian
\begin{equation}
\label{BlumeCapel}
H = -\beta \sum_{<xy>}  s_x s_y
  + D \sum_x s_x^2   \;\; ,   
\end{equation}
where the spin might assume the values $s_x \in \{-1, 0, 1 \}$.
$x=(x_0,x_1,x_2)$
denotes a site on the simple cubic lattice, where $x_i \in \{0,1,...,L_i-1\}$
and $<xy>$ denotes a pair of nearest neighbours on the lattice.
The inverse temperature is denoted by $\beta=1/k_B T$. The partition function
is given by $Z = \sum_{\{s\}} \exp(- H)$, where the sum runs over all spin
configurations. The parameter $D$ controls the
density of vacancies $s_x=0$. In the limit $D \rightarrow - \infty$
vacancies are completely suppressed and hence the spin-1/2 Ising
model is recovered. 

In  $d\ge 2$  dimensions the model undergoes a continuous phase transition
for $-\infty \le  D   < D_{tri} $ at a $\beta_c$ that depends on $D$, while
for $D > D_{tri}$ the model undergoes a first order phase transition,
where $D_{tri}=2.0313(4)$ for $d=3$, see ref. \cite{DeBl04}.

Numerically, using Monte Carlo simulations it has been shown that there
is a point $(D^*,\beta_c(D^*))$
on the line of second order phase transitions, where the amplitude
of leading corrections to scaling vanishes. 
We refer to the Blume-Capel model
at values of $D$ that are good numerical approximations of $D^*$ as
improved Blume-Capel model. For a more general discussion of improved models
see for example section 3.5 of \cite{myhabil} or section 2.3.1 of \cite{PeVi}.
In \cite{MHcritical} we
simulated the model at $D=0.655$ close to $\beta_c$ on lattices of a
linear size up to $L=360$. We obtained $\beta_c(0.655)=0.387721735(25)$
and $D^*=0.656(20)$.
The amplitude of leading corrections to scaling at $D=0.655$ is at
least by a factor of $30$ smaller than for the spin-1/2 Ising model.
Following eq.~(52) of ref. \cite{mytwopoint}, the amplitude  
of the second moment correlation length in the high temperature 
phase at $D=0.655$ is
\begin{eqnarray}
\label{xi0}
f_{2nd,+} &=&  0.2284(1) - 2.1 \times (\nu-0.629977)
                        + 500 \times (\beta_c - 0.387721735) \;\; \nonumber \\
&&  \mbox{using} \;\; t = \beta_c - \beta \;\;
 \mbox{as definition of the reduced temperature} .
\end{eqnarray}
Note that $\nu=0.629977$ is the estimate of the critical exponent of the 
correlation length given by ref. \cite{SiDu15}, which was the most 
accurate at the time. 
In the high temperature phase there is little difference between
$\xi_{2nd}$ and the exponential correlation length $\xi_{exp}$ which
is defined by the asymptotic decay of the two-point correlation function.
Following  \cite{pisaseries}:
\begin{equation}
\lim_{t\searrow 0} \frac{\xi_{exp}}{\xi_{2nd}} = 1.000200(3)
\;\;
\end{equation}
for the thermodynamic limit of the three-dimensional system.

\subsection{Definition of the interface free energy}
\label{defineFS}
Here we briefly recall a few basic definitions at the example of the 
Blume-Capel model. For a more detailed discussion see for example
\cite{BinderDG8,DG10,Cap} or
section 6 of ref. \cite{myhabil} and references therein.
Our starting point is the difference of the free energies of a system 
with an interface and one without. In order to force an interface
into the system, we consider so called anti-periodic boundary conditions.
These are implemented
by replacing in the reduced Hamiltonian, eq.~(\ref{BlumeCapel}),
the terms $\beta s_x s_y$ by
$-\beta s_x s_y$ for nearest neighbour pairs with $x_0=L_0-1$ and $y_0=0$
or vice versa.  
For the following discussion it is useful to introduce a reduced
Hamiltonian with a coupling $J_{<xy>,b}$ that depends on the link $<xy>$ and 
the type of the boundary conditions $b \in \{a,p\}$: 
\begin{equation}
\label{BlumeCapelG}
 H_b = - \beta \sum_{<xy>} J_{b,<xy>} s_x s_y
  + D \sum_x s_x^2  \;\; .
\end{equation}
In the case of periodic boundary conditions,
\begin{equation}
 J_{p,<xy>} = 1  \;\;\; \mbox{for all} <xy> \;.
\end{equation}
For anti-periodic boundary conditions
\begin{eqnarray}
 J_{a,<xy>} &=& - 1 \;\;\; \mbox{if} \;\; x_0=0 \;\; \mbox{and} \;\; y_0=L_0-1 \;\; \mbox{or vice versa.} 
\nonumber \\
 J_{a,<xy>} &=& 1  \;\;\; \mbox{else} .
\end{eqnarray}

Our first definition of the interface free energy is 
\begin{equation}
\label{fs1}
 F_s^{(1)} = - \ln(Z_a/Z_p) + \ln L_0 \;,
\end{equation}
where $\ln L_0$ takes into account the translational invariance. The 
partition function for the boundary condition $b$ is given by
\begin{equation}
 Z_b = \sum_{\{s\}} \exp(- H_b) \;\;.
\end{equation}
The definition~(\ref{fs1}) is motivated by the idea that for anti-periodic 
boundary conditions there is exactly one interface and no interface for 
periodic boundary conditions. A better approximation is given by
\begin{equation} 
\label{fs2}
 F_s^{(2)} = \ln L_0 - \ln \left( \frac{1}{2} \ln \frac{1+ Z_a/Z_p}{1- Z_a/Z_p} \right) \;,
\end{equation}
where it is assumed that for anti-periodic boundary conditions 
there is an odd number of interfaces, while for periodic ones there is 
an even number. It is assumed that these interfaces do not interact.
Note that $F_s^{(2)}$ in contrast to $F_s^{(1)}$ has a finite 
$L_0 \rightarrow \infty$ limit.

\subsection{Finite $L_0$ effects}
In this section we briefly review results obtained in  the literature.
For a more detailed discussion see section 6 of ref. \cite{myhabil} 
and references therein. 
The ratio of partition functions can be expressed in terms of eigenvalues
$\lambda$ of the transfer matrix $T$ in $0$-direction
\begin{equation}
\frac{Z_a}{Z_p} = \frac{\mbox{Tr} \; T^{L_0} P}{\mbox{Tr} \; T^{L_0}} =
\frac{\sum_i \left[ \lambda_{i,s}^{L_0} - \lambda_{i,a}^{L_0} \right] }
     {\sum_i \left[ \lambda_{i,s}^{L_0} + \lambda_{i,a}^{L_0} \right] } \;,
\end{equation}
where the matrix $P$ represents anti-periodic boundary conditions. 
Note that in the literature also the transfer matrix set up in a 
direction parallel to the interface has been considered, see for example
\cite{Rich93}.
The subscripts $s$ and $a$ stands for symmetric and anti-symmetric with respect 
to the spinflip $s_x \rightarrow -s_x$ for all $x$ on a slice of the lattice.
Let us assume that $i=0, 1, 2,...$ and $ \lambda_{i,s}$ and  $\lambda_{i,a}$
are decreasing with increasing $i$.  The symmetric eigenstates of the 
transfer matrix are eigenstates of $P$ with eigenvalue $1$ and 
the anti-symmetric ones are eigenstates of $P$ with eigenvalue $-1$. 
The tunneling correlation length is 
given by $\xi_t = -1/\ln(\lambda_{0,a}/\lambda_{0,s})$.   Here we consider 
the case that the tunneling correlation length $\xi_t$ is large compared with 
the bulk correlation length $\xi$.  Hence
\begin{equation}
 \lambda_{0s} > \lambda_{0a}  \gg  \lambda_{1s} > \lambda_{1a} , ...  \;.
\end{equation}
In the limit $L_1, L_2 \rightarrow \infty$, $\lambda_{is}$ and $\lambda_{ia}$
become degenerate.  The splitting decreases exponentially fast in $L_1, L_2$. 
Taking into account only the largest two eigenvalues one finds
\begin{equation}
\left( \frac{\lambda_{0,a}}{\lambda_{0,s}} \right)^{L_0} = \frac{Z_p-Z_a}{Z_p+Z_a} \;.
\end{equation}
Comparing with eq.~(\ref{fs2}) we get 
\begin{equation}
 F_s^{(2)} = \ln(2 \xi_t) \;\;,
\end{equation}
where for $L_1$, $L_2 \gg \xi_{exp}$ leading corrections are  
$O(\exp(-\L_0/\xi_{exp}))$. Note that the bulk correlation length is given by
\begin{equation}
\xi_{exp}=- \lim_{L_1,L_2 \rightarrow \infty} 1/\ln(\lambda_{1s}/\lambda_{0s})
   = - \lim_{L_1,L_2 \rightarrow \infty}  1/\ln(\lambda_{1a}/\lambda_{0a}) \;\;.
\end{equation}
A more accurate expression for the corrections would require precise 
knowledge of the splitting between $\lambda_{1,s}$ and $\lambda_{1,a}$ 
as a function of $L_1$ and $L_2$. 

The fact that the corrections vanish exponentially fast in $L_0$ enables
us to choose $L_0$ such that finite $L_0$ corrections can be completely 
ignored in the analysis of the data. Numerical experiments show that  $L_0$ 
taken to be a few times $L_1,L_2$ is sufficient to this end.
For details see section \ref{Snumerical} below.

\section{Predictions by the effective field theory}
\label{effectiveT}
Interfaces can be described by effective $d-1$-dimensional models, 
where $d$ is the dimension of the bulk system. 
In the context of statistical physics  such models are
called capillary wave models. For a review see e.g. ref. \cite{Cap}. 
In its simplest form it is  a massless Gaussian
theory, where the field corresponds to the transversal fluctuations
of the interface. 
 Note that in three dimensions,  by duality,
 interfaces correspond to strings in gauge
theories. Therefore effective theories describing such strings are 
in fact directly related to interfaces. In recent years there has been 
great progress in the understanding of the predictive power of such effective
models; see e.g. \cite{Aha,BrMe16} and refs. therein. 
It turns out that the Lorenz symmetry of the underlying gauge model,
or in our case the Galilean symmetry of the three-dimensional system, 
imposes constraints on the possible corrections to the free field theory.

In our study we are concerned with interfaces living on a torus with a
cross section of the size $L_1 L_2$. For the analysis of our data we need
the functional form of the free energy of the interfaces as a function
of $L_1$ and $L_2$. 

In the literature, the so called Nambu-Goto model is frequently discussed
as effective string model. 
Its action is proportional to the area of the interface.
The partition function of the Nambu-Goto model with periodic boundary 
conditions in both directions has been worked out in 
ref. \cite{billo06}. 
In the appendix of ref. \cite{billo06} the partition function is 
expanded in terms of powers of $1/(\sigma L_1 L_2)$.
For the free energy of the interface with periodic boundary conditions
in a three-dimensional system follows
\begin{equation}
\label{areaExp}
 F_s = \sigma L_1 L_2 + c_0 - \frac{1}{2} \ln \sigma -2 \ln \eta(i u)/\eta(i)  
    - f_1(u)  \frac{1}{ \sigma L_1 L_2} 
    - \tilde f_2(u)  \frac{1}{ (\sigma L_1 L_2)^2} + ... \;,
\label{Fsexpansion}
\end{equation}
where $ \tilde f_2(u) = f_2(u) - 0.5 f_1(u)^2$  and $u=L_1/L_2$ and 
$\eta$ is Dedekind's function
\begin{equation}
\eta(\tau) = q^{1/24} \prod_{n=1}^{\infty} (1-q^n) \;, \;\;\; q=\exp(2 \pi i \tau) \;. 
\end{equation}
Explicit expressions for $f_1(u)$ and $f_2(u)$ are given in eq.~(A.10) and 
(A.11) of ref. \cite{billo06}, respectively.   In our numerical study we 
consider the case $L_1=L_2$ throughout. One gets
\begin{equation}
f_1(1)  =  1/4     \;, \;\;\;  \tilde f_2(1)  =   -0.014107... \;\;.
\end{equation}
Note that $F_s = \sigma L_1 L_2 + \tilde c_0  -2 \ln \eta(i u)/\eta(i) $ 
is already predicted by the Gaussian interface model and does not rely
on the Galilean symmetry of the underlying system. The action of an effective 
interface
model could contained additional terms, such as a curvature term, that are 
not present in the 
Nambu-Goto model, altering the coefficients in eq.~(\ref{areaExp}). It turns
out that Galilean symmetry of the three-dimensional system ensures that terms 
up to $O(1/(\sigma L_1 L_2)^2)$ obtained from the Nambu-Goto model should
be valid.

For the model studied here, the Galilean symmetry is broken by the lattice 
and only restored in the critical limit.  We expect that the correction
exponent that is related to this restoration of symmetry is close to $2$
\cite{pisa97}. Furthermore, from the numerical results obtained in 
\cite{CaHaPa07} we conclude that for $\xi \approx 2$, the deviation of
$f_1(1)$ from its continuum value is of the order of $10 \%$.
We expect that the deviation in the case of the improved Blume-Capel model
has a similar amplitude as in the case of the Ising model.
Note that for real binary mixtures, Galilean symmetry should be present
at any temperature.

The constant $c_0$ in eq.~(\ref{Fsexpansion}) is not fixed by the effective
model. However Renormalization Group (RG)-theory predicts that 
\begin{equation}
C_0 = \lim_{\beta \searrow \beta_c} \; c_0(\beta) - \frac{1}{2} \ln[\sigma(\beta)] 
\label{C0RG}
\end{equation}
assumes a universal value. In ref. \cite{CaHaPa07} we obtained 
$C_0 = 0.3895(8)$ analysing our data obtained for the three-dimensional 
Ising model. A semiclassical calculation \cite{Muenster} gives 
$C_0 \approx 0.29$. See eqs.~(14,15) of ref. \cite{Klaus97}.

\section{The Monte Carlo algorithm}
Here we  essentially follow ref. \cite{CaHaPa07}, where the analogous problem was 
studied for the three-dimensional Ising model on the simple cubic lattice.
The main difference is that the interface energy $E_s$ is computed by using 
the variance reduced estimator discussed below.

We compute the interface free energy by
\begin{equation}
\label{integration}
F_s^{1}(\beta) = F_s^{1}(\beta_0) - \int_{\beta_0}^{\beta} \mbox{d} \tilde \beta  E_s(\tilde \beta)  \;\;,
\end{equation}
where the integration is performed numerically, using the trapezoidal rule. The starting
point of the integration, $F_s^{1}(\beta_0)$ is determined by using a variant 
of the boundary flip algorithm \cite{BF}.

The interface energy is defined as 
\begin{equation}
E_s =  E_a  -  E_p  \;,
\end{equation}
where 
\begin{equation}
 E_b = \langle \hat E_b \rangle\;\;\;,\;\;\;\; \hat E_b =\sum_{<xy>} J_{b,<xy>}  s_x s_y  \;.
\end{equation}
Note the unconventional sign that we take to be consistent with our previous work.
In \cite{CaHaPa07}, where we simulated the Ising model on the simple cubic
lattice, we performed independent simulations of systems with periodic and 
anti-periodic boundary conditions in order to determine $E_p$ and $E_a$. 

By using the exchange cluster algorithm \cite{ReMaCh98,ChMaRe98} we simulate
the systems with periodic and anti-periodic boundary conditions jointly.
The exchange cluster algorithm enables us to define a variance reduced
estimator of the difference $E_a-E_p$.
In the following, we recall the steps of the exchange cluster algorithm 
\cite{ReMaCh98,ChMaRe98,mysphere,myfilm}. Then we discuss the alignment 
of the configurations that is needed to get a considerable reduction of 
the variance. To get an ergodic update, the exchange cluster algorithm  has
to be supplemented by standard updates of the individual configurations.
Finally we summarize the complete update and measurement cycle. 

\subsection{The exchange cluster update}
\label{thealgorithm}
Let us briefly recall the basic properties of the exchange cluster update 
\cite{ReMaCh98,ChMaRe98}
at the example of our problem.
We simulate a system with periodic and a system with anti-periodic boundary conditions jointly.
The type of the boundary conditions is indicated by the first index of the field variable. Hence
$s_{p,x}$ and $s_{a,x}$ denote the spin at the site $x$ of the system with periodic and 
anti-periodic boundary conditions, respectively. 
The elementary step of the exchange cluster update is to swap the value of the spin
between the two systems:
\begin{eqnarray}
 s_{a,x}' = s_{p,x} \;\;\;,\;\; \nonumber
 s_{p,x}' = s_{a,x} \; .
\end{eqnarray}
This operation is performed for all sites within a cluster or for none. 
In ref. \cite{BrTa89} the cluster algorithm had been applied to the one
component $\phi^4$ model on the lattice. To this end embedded Ising variables
were introduced. The exchange cluster algorithm can be derived in a similar 
fashion. 
The swap of the spins can be written in terms of embedded Ising variables
$\sigma_x \in \{-1,1\}$:
\begin{eqnarray}
\label{embedI}
s_{a,x}' &=& \frac{1 +\sigma_x}{2} s_{a,x} + \frac{1 -\sigma_x}{2} s_{p,x} \;,  \nonumber \\
s_{p,x}' &=& \frac{1 +\sigma_x}{2} s_{p,x} + \frac{1 -\sigma_x}{2} s_{a,x} \;.
\end{eqnarray}
For $\sigma_x=-1$ the exchange is performed, while for $\sigma_x=1$ the old 
values are kept. Plugging eq.~(\ref{embedI}) into the reduced 
Hamiltonian $H(\{s_a'\},\{s_p'\})=H_a(\{s_a'\}) + H_p(\{s_p'\})$, 
eq.~(\ref{BlumeCapelG}), one reads off the 
coupling constants $\beta_{embed,<xy>}$ for the embedded Ising variables. 
The clusters are defined by frozen links. Frozen links are those links 
that are not deleted. The probability to delete a link $<x,y>$ is given by
\cite{BrTa89}:
\begin{equation}
\label{pdsimple}
 p_{d,<xy>} = \mbox{min} [1, \exp(-2 \beta_{embed,<xy>} )]  \;,
\end{equation}
where following eq.~(24) of  \cite{myfilm}
\begin{equation}
\beta_{embed,<xy>} =\beta \frac{J_{p,<xy>} + J_{a,<xy>}}{4} ( s_{p,x} - s_{a,x} )  ( s_{p,y} - s_{a,y} ) \;.
\end{equation}
Hence 
\begin{equation}
\beta_{embed,<xy>\in B} =0 \;\;\;\mbox{and} \;\; 
\beta_{embed,<xy>\notin B} = \frac{\beta}{2} ( s_{p,x} - s_{a,x} )  ( s_{p,y} - s_{a,y} ) \;,
\end{equation}
where $B$ denotes the set of all pairs of nearest neighbours $<xy>$ with $x_0=0$ and $y_0=L_0-1$, or
vice versa.
As discussed in section IV of \cite{myfilm}, for $J_{p,<xy>} \ne J_{a,<xy>}$, 
an external field arises, eq.~(25) of \cite{myfilm}:
\begin{equation}
\label{embedgeneral}
 h_{embed,x,<xy>} = \beta \frac{J_{p,<xy>} - J_{a,<xy>}}{4}  (s_{p,x} - s_{a,x}) ( s_{p,y} + s_{a,y} )
\;,
\end{equation}
where the indices $x,<xy>$ refer to the fact that the field at the site $x$ arises from the 
interactions on the link $<xy>$. 
This means in our case
\begin{equation}
\label{embedh}
 h_{x,<xy>,embed} = \frac{\beta}{2}  (s_{p,x} - s_{a,x}) ( s_{p,y} + s_{a,y} )
\end{equation}
for  $x_0=L_0-1$ and $y_0=0$ or vice versa. In total the embedded external field is
$h_{x,embed} = \sum_{y.nn.x} h_{x,<xy>,embed}$, where $y.nn.x$ means that $y$ is 
a nearest neighbour of $x$. Hence there might be  a non-vanishing external field at $x_0=0$
and $x_0=L_0-1$, while it vanishes at all other sites.  For a given 
decomposition into clusters there is huge freedom in the selection of 
clusters, where the exchange of spins is performed.  As long as for a given
decomposition into clusters, the probability to undo the exchange is the 
same as the exchange itself, detailed balance is satisfied.
Similar to refs. \cite{mysphere,myfilm} it seems optimal to exchange as many spins as possible
between the two systems. 
The exchange of spins is only hindered by the effective external field $h_{x,embed}$. Hence we 
only compute those clusters, which are frozen by the external field. To this end,
we first run through the planes given by $x_0=0$ and $x_0=L_0-1$. A site is
frozen with the probability
$ p_{f,h} = 1-p_{d,h} $,
where 
\begin{equation}
p_{d,h} = \mbox{min} [1, \exp(-2 h_{x,embed} )] \;\;.
\end{equation}
After running through these two planes, freezing sites, we construct
all clusters that contain sites that are frozen due to the external field. To this end, 
the probability~(\ref{pdsimple}) is used.
Then the spins in these clusters remain unchanged, while the spins at all other sites 
are exchanged between the two systems. 

\subsection{The variance reduced estimator of the difference of the internal energies}
Following eq.~(30) of \cite{myfilm} the variance reduced estimator of the
difference is given by 
\begin{equation}
\label{improvedE}
\hat E_{s,imp} = \frac{1}{2} \left[(\hat E_a -\hat E_p') + (\hat E_a' -\hat E_p) \right] \;,
\end{equation}
where $\hat E_a$ and $\hat E_p$ are the standard estimators evaluated before and 
$\hat E_a'$ and $\hat E_p'$ after the exchange cluster update. Since the spin is 
exchanged for all sites that do not belong to the frozen clusters, we get 
an exact cancellation for all nearest neighbour pairs $<xy> \notin B$, where
both $x$ and $y$ do not belong to the frozen exchange clusters.

In our program, we evaluate the contributions from pairs of nearest neighbour
sites $<xy> \in B$, by implementing eq.~(\ref{improvedE}) directly:
\begin{equation}
\label{improvedE1}
\Delta \hat E_{imp,1} = - \frac{1}{2} \sum_{<xy> \in B} 
s_{a,x} s_{a,y}   + s_{p,x} s_{p,y} +
s_{a,x}' s_{a,y}' + s_{p,x}' s_{p,y}' \;.
\end{equation}
The contribution from nearest neighbour pairs where both sites belong to the 
frozen clusters and $<xy>$ is not in the boundary:
\begin{equation}
\Delta \hat E_{imp,2} =  \sum_{<xy>\notin B, x \in C, y \in C} s_{a,x} s_{a,y} - 
                                      s_{p,x} s_{p,y} \;.
\end{equation}
And finally the contribution, where $<xy>$ is not in the boundary, the site $x$
is in the frozen clusters, but $y$ is not
\begin{equation}
\label{improvedE3}
\Delta \hat E_{imp,3} = \frac{1}{2} \sum_{<xy> \notin B,x \in C, y \notin C}
s_{a,x} (s_{a,y}+s_{p,y}) - s_{p,x} (s_{a,y}+s_{p,y}) \;.
\end{equation}
In total
\begin{equation}
\label{improvedEV}
\hat E_{s,imp} = \Delta \hat E_{imp,1} + \Delta \hat E_{imp,2} + \Delta \hat E_{imp,3} \;\;.
\end{equation}
We note that the effort to compute $\hat E_{s,imp}$ is essentially proportional
to the total volume of the frozen exchange clusters. In the following we shall
denote the collection of frozen exchange clusters simply by exchange cluster. 

\subsection{Aligning the configurations}
In refs. \cite{myfilm,mytwopoint} we learnt that in the case of spontaneous symmetry
breaking, it is important to align the magnetisation of the two systems that are
simulated. This way, the frozen exchange clusters remain small compared with the volume of the 
system and the variance reduction is effective. In the case of the problem studied here,
two steps are needed to this end. First we exploit the translational symmetry of the system
with anti-periodic boundary conditions in $0$-direction to shift the interface between 
the phases to the boundary. This way, typical configurations show a unique magnetisation.

A backward shift by $i_s \in \{0,1,2,...,L_0-1\}$ is performed in the following way:
For anti-periodic boundary conditions, if $x_0+i_s < L_0$
\begin{equation}
\label{sh1}
s_{a,x_0,x_1,x_2}'=  s_{a,x_0+i_s,x_1,x_2}
\end{equation}
and else
\begin{equation}
\label{sh2}
s_{a,x_0,x_1,x_2}' = - s_{a,x_0+i_s-L_0,x_1,x_2}  \;.
\end{equation}
For periodic boundary conditions, if $x_0+i_s < L_0$
\begin{equation}
\label{sh3}
s_{p,x_0,x_1,x_2}' =  s_{p,x_0+i_s,x_1,x_2}
\end{equation}
and else
\begin{equation}
\label{sh4}
s_{p,x_0,x_1,x_2}' = s_{p,x_0+i_s-L_0,x_1,x_2}  \;.
\end{equation}

To get a correct algorithm, we need a characterization of the position of 
the interface that 
is not altered by the exchange cluster update. To this end, we compute
\begin{equation}
 P(x_0) = \sum_{x_1,x_2}  s_{a,x} s_{p,x} \;\;,
\end{equation}
which remains unchanged by the exchange of spins between the two systems.
For a similar construction see eq.~(8) of ref. \cite{mytwopoint}.

At the centre of the interface, we expect that the number of spins 
with $s_{a,x}=-1$ and with $s_{a,x}=1$ is roughly the same. 
Therefore the position of the 
interface should be given by the position $x_{0,min}$ of the minimum 
of $|P(x_0)|$. If the minimum of $|P(x_0)|$ is degenerate, we pick one randomly.
In order to shift the interface to $x_0=0$, we choose $i_s=x_{0,min}$. 
Furthermore,  we like to have
the same sign of the overall magnetisation for both systems.  Therefore, after
shifting, with newly computed $P(x_0)$, 
if $\sum_{x_0} P(x_0) < 0$ we multiply $s_{p,x}$ by $-1$ for all
$x$. In case $\sum_{x_0} P(x_0)=0$  we perform the operation with
probability $1/2$. After this alignment is done, the exchange cluster
update as discussed above is performed along with the measurement of the 
variance reduced estimator of the energy difference. Next we undo the 
alignment:
With probability $1/2$
all spins of the system with periodic boundary conditions are multiplied by
$-1$.
Then random shifts, eq.~(\ref{sh1},\ref{sh2},\ref{sh3},\ref{sh4}),
by $i_{s,a}$ and $i_{s,p}$ for anti-periodic and periodic boundary 
conditions, respectively, are performed. The values of $i_{s,a}$ 
and $i_{s,p}$ are selected in $\{0,1,2,...,L_0-1\}$ with equal probability. 

\subsection{The complete update cycle}
Since the exchange cluster update is not ergodic, it is supplemented by
standard updates of the individual systems. To this end we use the local 
heat-bath algorithm, 
the local Todo-Suwa \cite{ToSu13,Gutsch} algorithm and standard single cluster 
updates \cite{Wolff}, where ergodicity is provided by the local heat-bath 
algorithm.
An update cycle is composed by one sweep with the local 
heat bath algorithm for both systems, $N_{clu}$ single cluster updates of the system with 
periodic boundary conditions only and one sweep with the local Todo-Suwa algorithm 
for both systems. Finally $N_{ex}$ exchange cluster updates are performed along
with the alignment and the shifts discussed above.

We chose $N_{clu}$ such that 
$N_{clu}$ times the average cluster size is roughly equal to the volume of the 
lattice. We update only the system with periodic boundary conditions by using
the single cluster algorithm, since here in contrast to the anti-periodic 
boundary conditions, the introduction of an auxiliary array that indicates 
whether a site belongs to the cluster is not required.  The number $N_{ex}$ 
is chosen as odd number. This way configurations of the systems with periodic
and anti-periodic boundary conditions are effectively swapped,  avoiding 
large autocorrelation times for the system with anti-periodic boundary 
conditions,
where no single cluster updates are performed.  Note that it turns out that 
the clusters that have to be constructed for the exchange cluster update take
on average only a small fraction of the volume of the system. Therefore, similar
to $N_{clu}$ we choose $N_{ex}$ such that $N_{ex}$ times the average total 
cluster size equals roughly the volume of the lattice.

Note that in the actual program, to save CPU-time, the spin values are 
exchanged only for the 
frozen clusters, while at the same time the type of the boundary conditions 
is swapped.

Let us summarize the steps of one update cycle by using a piece of pseudo-C 
code:

\begin{verbatim}

sweep with the local heat bath algorithm for both systems;
for(iclu=0;iclu<nclu;iclu++)
  {
  single cluster update of the system with periodic boundary conditions;
  }
sweep with the local Todo-Suwa algorithm for both systems;
for(iex=0;iex<nex;iex++)
  {
  align configuations;
  exchange cluster update with measurement of variance reduced energy
  difference;
  unalign configuations;
  }

\end{verbatim}

For a more formal discussion of the algorithm it is useful to write the 
updates as matrices $P_{AL}$ that act on probability distributions. 
The subscript $AL$ indicates the algorithm that is used. The indices of the matrix
are given by the configuations. One update cycle is represented by the matrix
\begin{equation}
\label{Pcycle}
 P_{cycle} =  P_{U,EX,A}^{N_{ex}} \; P_{TS} \; P_{SC}^{N_{clu}} \; P_{HB} \;\;,
\end{equation}
where $HB$ denotes a sweep over both systems using the heat-bath
algorithm, $SC$ a single cluster update of the system with periodic
boundary conditions, $TS$ a sweep over both systems using the Todo-Suwa
algorithm, and $U,EX,A$ the exchange cluster update along with the
alignment and unalignment of the configuations for periodic and anti-periodic 
boundary conditions. Here we follow the convention
that the matrices act on vectors on the right. A correct algorithm 
should be ergodic and should satisfy stability
\begin{equation}
\label{stability}  
w = P_{cycle} w \;,
\end{equation}
where $w$ denotes the distribution that we intend to generate, which is in 
our case the Boltzmann distribution $w=\exp(-H_a-H_p)/Z$. Eq.~(\ref{stability})
is satisfied if stability is satisfield for each of the factors of $P_{cycle}$.
For the Todo-Suwa algorithm, the single cluster algorithm and the 
heat-bath algorithm this has been shown in the literature.   The alignment 
of the configurations modifies the Boltzmann distribution by introducing a
constraint. Let us denote this distribution by $\tilde w$.  We constructed 
the constaint such that it is kept by the exchange cluster update. 
Furthermore, the exchange cluster update inherits detailed balance 
from the cluster update of the embedded Ising model \cite{BrTa89}. Hence
$\tilde w = P_{EX} \tilde w$.  Finally, the unalignment restores the 
Boltzmann distribution $w$ from the Boltzmann distribution with constraint 
$\tilde w$.  Hence $w=P_{U,EX,A} w$.
\section{Numerical results}
\label{Snumerical}
We performed simulations
with the boundary flip algorithm \cite{BF} to get the starting value for the 
integration~(\ref{integration}). The results for $F_s^{(1)}(\beta_0)$, 
eq.~(\ref{fs1}), are summarized in table \ref{Fstart}. 
Here we do not go further into the details
of the simulations, since they are very similar to those of
ref. \cite{CaHaPa07}.
We just note that the boundary flip algorithm becomes inefficient in the 
limit $Z_a/Z_p \rightarrow \infty$. Practically one is limited to 
$\sqrt{\sigma} L \lessapprox 4$ for $L=L_1=L_2$. 
To reach larger values of $\sqrt{\sigma} L$, 
we perform the numerical integration of $E_s$ over $\beta$, 
eq.~(\ref{integration}).

\begin{table}
\caption{\sl \label{Fstart}
Numerical results for $F^{(1)}(\beta_0)$ obtained for lattices 
with $L_1=L_2=L$ by using the boundary flip algorithm \cite{BF}. 
}
\begin{center}
\begin{tabular}{lrrc}
\hline
\mc{1}{c}{$\beta_0$} & \mc{1}{c}{$L$} &   \mc{1}{c}{$L_0$}  &
 $F^{(1)}$ \\
\hline
0.391  &  32  & 32 & 8.49222(39) \\
0.391  &  32  & 64 & 8.54138(39) \\
0.391  &  32  &128 & 8.54183(55) \\
0.3885 &  64  & 64 & 7.44256(60)  \\
0.3885 &  64  &128 & 7.50640(40)  \\
0.3885 &  64  &256 & 7.5101(13) \\
0.388  & 128  &128 & 8.4422(25)  \\
0.388  & 128  &256 & 8.50764(88)  \\
0.388  & 128  &512 & 8.5109(15)  \\
0.38776& 256  &256 & 6.9731(19)  \\ 
0.38776& 256  &512 & 7.0493(18)  \\  
\hline
\end{tabular}
\end{center}
\end{table}

\subsection{Computing $E_s$ by using  the exchange cluster algorithm}
We implemented the code in standard C and used the SIMD-oriented Fast Mersenne
Twister algorithm \cite{twister} as random number generator.  As check of the 
code, we performed high statistics simulations for $L=L_1=L_2=2$ and $L_0=3$ 
and $4$. For comparison we computed the observables exactly, up to rounding 
errors, by performing the sum over all configurations.
Our simulation program passed this benchmark.

We performed a large number of simulations with the  exchange cluster algorithm
for lattices of the size $L=L_1=L_2$ and $L_0=L$, $2 L$, or $4 L$. In 
particular we considered $L=32$, $64$, $128$, and $256$ in the range
$0.391 \le \beta \le 0.44$, 
$0.3885 \le \beta \le 0.44$, $0.388 \le \beta \le 0.4$, and
$0.38776 \le \beta \le 0.394$, respectively.  Details are summarized in 
table \ref{Simulations}. In certain intervals $[\beta_a,\beta_f]$ we
simulated at sampling points $\beta_i$ that are separated by 
$\Delta \beta=\beta_{i+1}-\beta_i$.  For example for $L=L_0=256$ we
simulated at 581 different values of $\beta$.   Note that the intervals
and the step size $\Delta \beta$ are chosen such that the systematic error
of the numerical integration 
is about one order of magnitude smaller than the statistical error.
The systematic error of the
numerical integration was estimated by thinning out the sampling points.
For $L=32$ we performed $10^6$ update cycles after equilibration  for each
value of $\beta$. For larger $L$ we performed fewer updates. For example 
for $L=L_0=256$ for $\beta \le 0.391$ we performed $10^5$ update cycles 
after equilibration. For $0.391 < \beta \le 0.393$ we performed $4 \times 10^4$
update cycles and for $0.393 < \beta \le 0.394$ we performed $2 \times 10^4$
update cycles. The simulations for $L=L_0=256$ and $10^5$ update cycles took 
about 6 days on a single core of a Xeon(R) E5-2660  CPU each.

\begin{table}
\caption{\sl \label{Simulations}
List of the simulations with the exchange cluster algorithm.
In the first column we give $L=L_1=L_2$, in the second column the 
values of $L_0$ that have been simulated. In the third and fourth 
column we give the start $\beta_a$ and end point 
 $\beta_f $ of the interval in the inverse temperature that is 
considered.  Finally in the last column, we give the step size 
$\Delta \beta$   that is used in the interval.
}
\begin{center}
\begin{tabular}{rclll}
\hline
$L$ & $L_0$ & $\beta_a$ & $\beta_f $ & $\Delta \beta$ \\
\hline
32 & 32,64,128 & 0.391 & 0.394 & 0.00002 \\
32 & 32,64,128 & 0.394 & 0.396 & 0.00005 \\
32 & 32,64,128 & 0.396 & 0.4   & 0.0001  \\
32 & 32,64   & 0.4   & 0.42  & 0.0005  \\
32 & 32 &  0.42  & 0.44  & 0.001 \\
\hline
64  & 64,128,256 &  0.3885 & 0.391 & 0.00001\\
64  & 64,256 &  0.391  & 0.394 & 0.00002 \\
64  & 128 &  0.391  & 0.392 & 0.00002 \\
64  & 64 &  0.394  & 0.397 & 0.00005 \\
64  & 64 &  0.397  & 0.4   & 0.0001 \\
64  & 64 &  0.4    & 0.42  & 0.0005 \\
64  & 64 &  0.42   & 0.44  & 0.001  \\
\hline
128 & 128 & 0.388 &0.391 & 0.00001 \\
128 & 256 & 0.388 & 0.3885 & 0.000005\\
128 & 256 & 0.3885 & 0.39 & 0.00001\\
128 & 512 & 0.388 & 0.38955 & 0.00001\\
128 & 128 & 0.391 &0.394 & 0.00002 \\
128 & 128 & 0.394 &0.3975& 0.00005 \\
128 & 128 &  0.3975&0.4   & 0.0001\\
\hline
256 & 256 & 0.38776 & 0.38780  & 0.000001 \\
256 & 256 & 0.38780 & 0.38790  & 0.000002 \\
256 & 256 & 0.38790 & 0.38820  & 0.000005 \\
256 & 256 & 0.38820 & 0.39100  & 0.00001 \\
256 & 256 & 0.39100 & 0.39400  & 0.00002 \\
256 & 512 & 0.38776 & 0.38820  & 0.00001 \\
\hline
\end{tabular}
\end{center}
\end{table}

First we investigated the performance 
of the exchange cluster update and the variance reduced estimator associated 
with it. Then we analysed our numerical results obtained for the interface 
energy and free energy.

\subsubsection{Average size of the exchange cluster per area}
\begin{figure}
\begin{center}
\includegraphics[width=14.5cm]{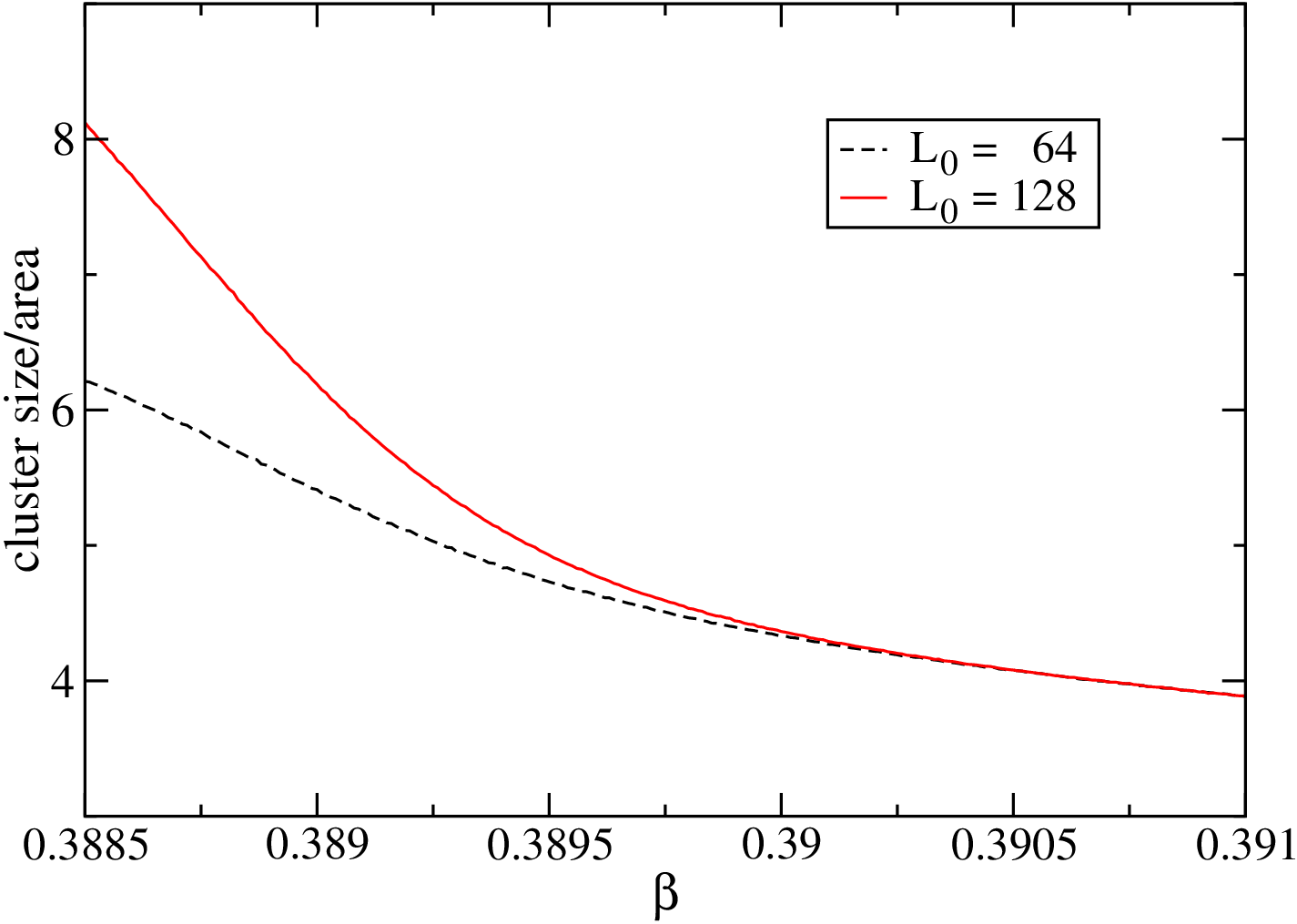}
\caption{\label{plot64c}
We plot the size of the exchange cluster per area  for $L=64$ and the two 
choices $L_0=64$ and $128$ as a function of $\beta$. 
}
\end{center}
\end{figure}

In figure \ref{plot64c} we plot the average size of the exchange cluster per area $C_{ex}$ for 
$L=64$ and the two lengths $L_0=64$ and $128$. Here area is $L_1 L_2$ and the 
size of the cluster is the number of sites contained in it.
For $\beta \gtrapprox 0.3905$ 
the sizes for $L_0=64$ and $128$ can not be discriminated at the level of our
statistical accuracy. For smaller values of $\beta$, the cluster size for 
$L_0=128$ is larger than that for $L_0=64$. 
For $L=32$ and the lengths $L_0=32$ and $64$ we find that for 
$\beta \gtrapprox 0.396$, the cluster sizes for $L_0=32$ and $64$ can not 
be discriminated. Also here, for smaller values of $\beta$, the cluster size
is larger for the larger $L_0$.  In the case of $L=128$ and $L_0=128$ we generated
only data for $\beta \ge 0.389$. For these values of $\beta$ we find that
the cluster sizes agree at the level of our accuracy for $L_0=128$ and $256$. 
For $L=256$ we find that for $\beta \gtrapprox 0.38805$ 
the sizes for $L_0=256$ and $512$ can not be discriminated at the level of our
statistical accuracy. Again, for smaller values of $\beta$, 
the cluster size is larger for the larger $L_0$. 
These threshold values of 
$\beta$ correspond to $L/\xi_{exp} \approx 12.8, 13.0,$ and $13.6$ for 
$L=32$, $64$, and $256$, respectively. 

We interpret these findings as follows: For sufficiently low values of $\beta$, 
the probability to have more than one interface is negligible. If there is only
one interface, the exchange cluster contains only sites in the neighbourhood 
of the boundary. The size of the exchange cluster is governed by the interface.
Hence as soon as $L_0$ is large compared with the width of the interface, there
is no dependence of the size of the exchange cluster on $L_0$.  

Next let us study the dependence of the cluster size per area on $L$.
For example at $\beta=0.396$, where we see virtually no dependence on $L_0$,
we find $2.6751(14)$, $2.9827(16)$,
$3.2523(36)$, and $3.4935(46)$ for $L=32$, $64$, $128$, and $256$ respectively. 
Note that the simulation for $L=256$ at $\beta=0.396$ was performed mainly to 
get cluster size.
The behaviour of the cluster size is roughly consistent 
with a logarithmic growth in $L$.  Using the Ansatz
\begin{equation}
 C_{ex}^2 = c + a \ln L
\end{equation}
we get $c= -1.37(4)$, $a=2.463(11)$ and $\chi^2/$d.o.f$=7.18$ taking all
four values of $L$ into account. Skipping $L=32$ we get 
$c=-1.07(9)$, $a=2.397(21)$, and $\chi^2/$d.o.f$=0.88$. A similar fit for 
$C_{ex}$ itself produces much larger values of  $\chi^2/$d.o.f..

\subsubsection{Reduction of the statistical error}

\begin{figure}
\begin{center}
\includegraphics[width=14.5cm]{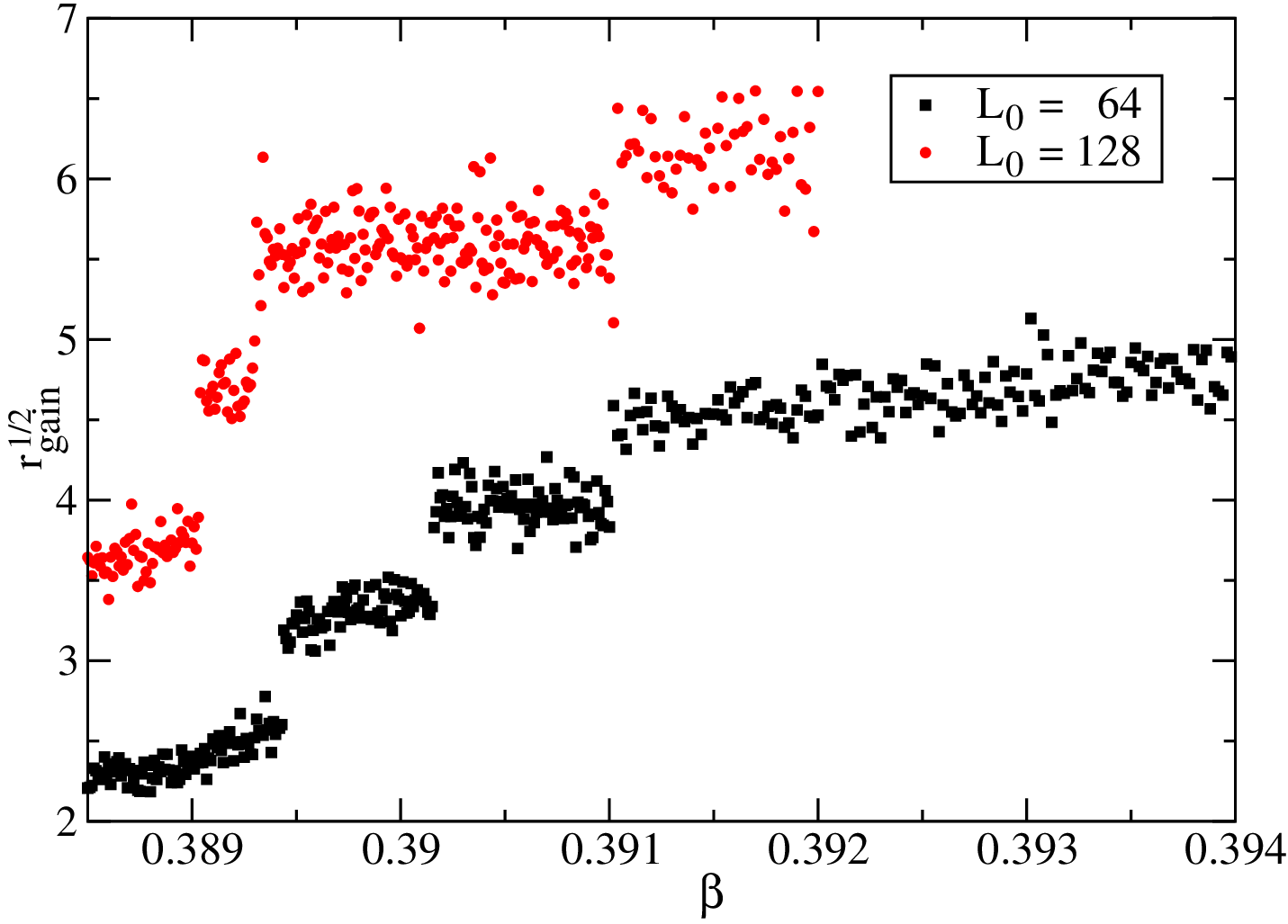}
\caption{\label{rgain}
We plot $r_{gain}^{1/2}$, eq.~(\ref{rga}), for $L=64$ and $L_0=64$ and $128$
as a function of $\beta$. 
}
\end{center}
\end{figure}

We computed the naive statistical error of the difference of the internal 
energies as
\begin{equation}
 \epsilon^2(E_s) = \epsilon^2(E_a) + \epsilon^2(E_p) \; ,
\end{equation}
where we ignore correlations between the two systems, which are caused by the 
exchange cluster update. This should resemble quite well the situation of 
independent simulations for periodic and anti-periodic boundary conditions.
The improvement that we get by using the cluster exchange algorithm and 
the variance reduced estimator associated with it is characterized by
\begin{equation}
\label{rga} 
r_{gain} = \epsilon^2(E_s)/\epsilon^2(E_{s,imp}) \;\;. 
\end{equation}
In figure \ref{rgain}  we plot $r_{gain}^{1/2}$ 
 for $L=64$ and $L_0=64$ and $128$. For other lattice sizes $L,L_0$ we get
similar results.
We find that for fixed parameters of the algorithm, the gain
increases with increasing $\beta$ and $L_0$.  The steps in $r_{gain}^{1/2}$,
plotted as a function of $\beta$, are due to a change of $N_{ex}$. 
For $L_0=64$  
we use $N_{ex}=3$ up to $\beta=0.38943$, $N_{ex}=5$ from $\beta=0.38944$ up to
$0.39015$, $N_{ex}=7$ from $\beta=0.39016$ up to $\beta=0.391$ and 
$N_{ex}=15$ from $\beta=0.39102$ up to $0.44$. For $L_0=128$, 
we use $N_{ex}=3$ up to $\beta=0.38903$, $N_{ex}=5$ from $\beta=0.38904$ 
up to $0.3893$, $N_{ex}=7$ from $\beta=0.38931$ up to $0.391$, $N_{ex}=9$
for $\beta=0.39102$ and $N_{ex}=15$ for $\beta=0.39104$ up to $0.3920$. 
Note that we made no effort to fine tune the parameter $N_{ex}$ of the 
algorithm. It is chosen such that  $N_{ex}$ times the average size of the 
exchange cluster is roughly equal to the lattice size.
 In our simulations we find a gain $r_{gain}$, depending on $\beta$, $L$, and
$L_0$ that ranges from a factor of $\approx 4$ up to $\approx 70$. 

In a preliminary stage of our study, we had implemented the 
exchange cluster algorithm without the alignment of the configurations
discussed in section \ref{thealgorithm}. In this case we see a much 
larger size of the exchange cluster. Furthermore the reduction 
of the variance is moderate.

\subsection{Finite $L_0$ effects}
\begin{figure}
\begin{center}
\includegraphics[width=14.5cm]{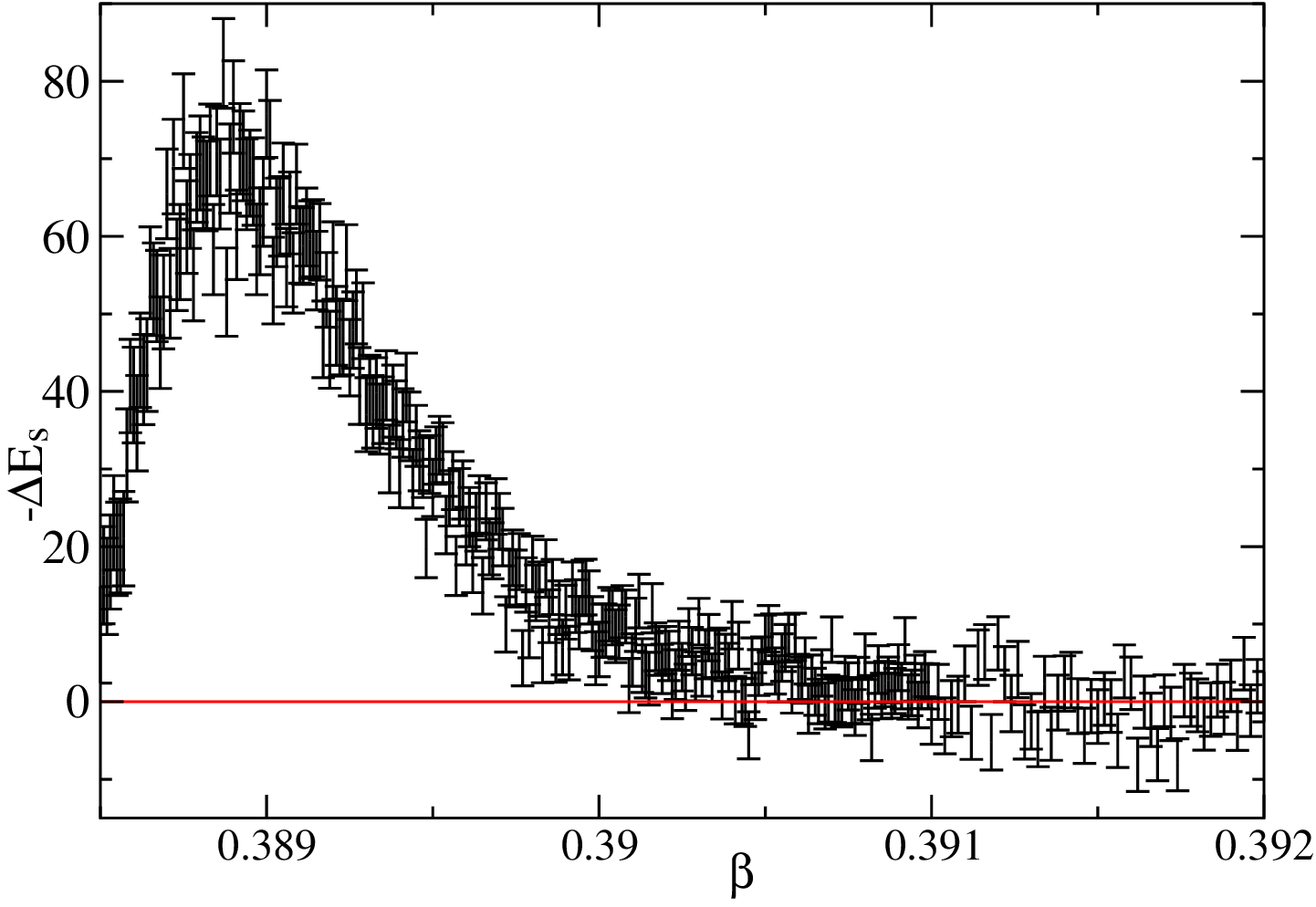}
\caption{\label{DeltaEs}
We plot minus the difference of the interface energy for $L=64$ between 
$L_0=64$ and $128$. The solid red line simply indicates zero.
 For a discussion see the text.  
}
\end{center}
\end{figure}

Next we investigated finite $L_0$ effects
in $E_s$ and $F_s$. In figure \ref{DeltaEs} we plot minus the difference 
$\Delta E_s = E_s(L=64,L_0=64) -  E_s(L=64,L_0=128)$. We find that
$\Delta E_s $ vanishes within the statistical errors for
$\beta \gtrapprox 0.3907$. In the following, to be on the safe side,
we shall assume that for $L_0=L=64$ finite $L_0$ effects can be
ignored for $\beta \gtrapprox 0.3915$. 

\begin{figure}
\begin{center}
\includegraphics[width=14.5cm]{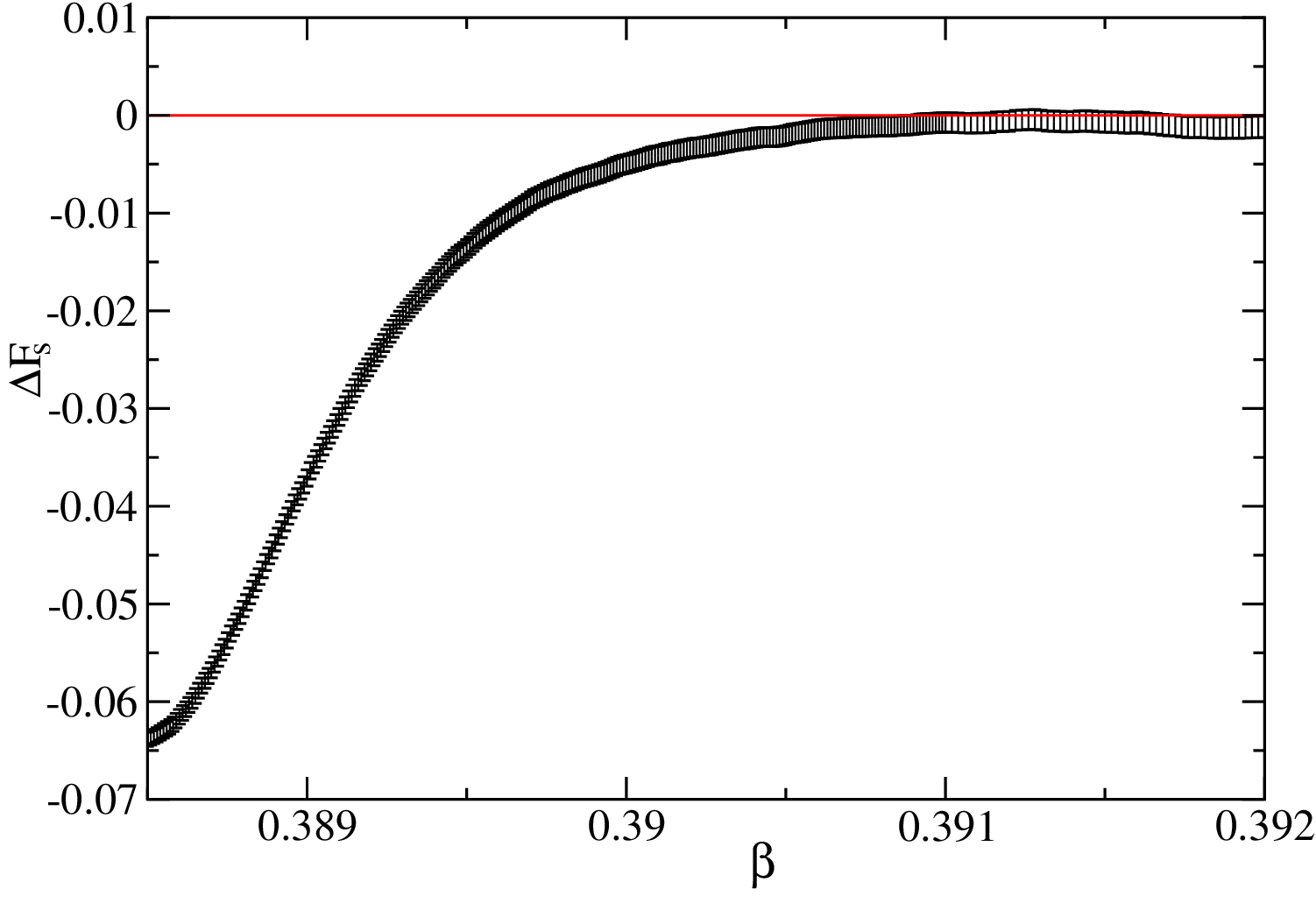}
\caption{\label{DeltaFs}
We plot the difference of the interface free energy for $L=64$ between
$L_0=64$ and $128$. The solid red line simply indicates zero.
 For a discussion see the text.
}
\end{center}
\end{figure}

In figure  \ref{DeltaFs} we show the corresponding difference 
$\Delta F_s = F_s(L=64,L_0=64) -  F_s(L=64,L_0=128)$.
Similar to the finding above, we find that $\Delta F_s$ vanishes within the 
statistical errors for $\beta \gtrapprox 0.3907$. Here we should note that  
results for different values of $\beta$ are statistically correlated due
to the fact that $F_s$ is obtained by integrating $E_s$.  In the following
we assume that at our level of statistical accuracy, for
$L_0=L=64$ finite $L_0$ effects in $F_s$ can be safely ignored for 
$\beta \gtrapprox 0.3915$. Translating this into a dimensionless ratio, 
we get that for $L_0=L$ finite $L_0$ effects in $F_s$ can be safely ignored
if $L/\xi_{exp} \gtrapprox  16$. 
Checking our numerical results for $L=32$, $128$, and 
$256$, we confirm this finding.
Performing a similar analysis, we conclude that for $L_0=2 L$ finite 
$L_0$ effects in $F_s^{(2)}$ can be ignored if $L/\xi_{exp} \gtrapprox 6$.
Note that all data used below satisfy these requirements.  
For our final estimates of $F_s$ obtained by integrating  $E_s$,
eq.~(\ref{integration}), 
we lowered the value of $L_0$ by a factor of two at certain values of $\beta$,
where the difference between $E_s(L,L_0,\beta)$ and $E_s(L,2 L_0,\beta)$
is negligible.  For example for $L=32$, we started the integration at 
$\beta=0.391$ with $L_0=128$. For $0.393 < \beta \le 0.4$ we used $E_s(32,64)$
and then for $0.4 < \beta \le 0.44$ we used instead $E_s(32,32)$. On the 
other hand for $L=256$ we used $L_0=256$ throughout.  The particular choice 
for each value of $L$ is related to the accuracy of data that we had generated.

\subsection{Critical behaviour of the interface energy $E_s$}
The interface free energies that are obtained from integrating the interface
energies are, by construction, statistically correlated. In order to avoid this
complication we analyse directly the interface energies.

We start from the Ansatz
\begin{equation}
\label{ansatzfs}
 F_s = \sigma(\beta) L^2  + c_0(\beta) - \frac{1}{4} \frac{1}{\sigma(\beta) L^2} 
\end{equation}
for the interface free energy, where 
\begin{equation}
\label{sigmaansatz}
 \sigma(\beta)  = \sigma_0 t_m^{2 \nu} \;\; [1 +a t_m^{\theta} +  b t_m+ ... ] \;\;,
\end{equation}
where we expect that $|a|$ is small, since we study an improved model. 
For convenience we have introduced $t_m=-t = \beta - \beta_c$ here.
Following RG-theory
the constant behaves as
\begin{equation}
\label{cansatz}
 c_0(\beta) =  c + \ln \xi + ... = \tilde c - \nu \ln t_m  + \bar c t_m^{\theta} + \bar d t_m + ...
\end{equation}
where again $|\bar c|$ is expected to be small.

Taking the derivative of eq.~(\ref{ansatzfs}) with respect to $\beta$, we arrive at
\begin{equation}
\label{finalansatz}
 E_s = 2 \nu \sigma_0  t_m^{2 \nu -1} [1 + \tilde a t_m^{\theta} +  \tilde b t_m ]  L^2
      -\nu t_m^{-1} +   
       \bar d  
  + \frac{1}{4 L^2} \frac{2 \nu}{\sigma_0} t_m^{-2 \nu -1}
\frac{1 + \tilde a t_m^{\theta} +  \tilde b t_m }{(1 +a t_m^{\theta} +  b t_m)^2} \;\;,
\end{equation}
where we ignored corrections that are represented by $...$ in eqs.~(\ref{sigmaansatz},\ref{cansatz}).
Furthermore, we skipped the term $\bar c \theta  t_m^{\theta-1}$ which should be negligible for the 
improved model.
We define
\begin{equation}
\tilde a =a \left(1+\frac{\theta}{2 \nu} \right)  \;\;\;, \;\;\;
\tilde b =b \left(1+\frac{1}{2 \nu} \right)
\end{equation}
to keep eq.~(\ref{finalansatz}) readable.

\subsubsection{Numerical results}
We have fitted our numerical data for $L=32$, $64$, $128$, and $256$ using the 
Ansatz~(\ref{finalansatz}).
We fixed $\nu=0.6299709$, $\omega=0.82968$ and $\beta_c=0.387721735$.
In order to keep finite $L$ effects small, we took only data with 
$\sqrt{\sigma} L  \gtrapprox 6$ into account. In order to estimate systematic 
errors due to subleading corrections that are not included in the Ansatz and 
due to deviations of the coefficient of $1/(\sigma(\beta) L^2)$ from $-1/4$, 
we varied the 
range of parameters that are included into the fit. For example for 
$\sqrt{\sigma} L  \gtrapprox 6$ and $\beta \le 0.392$ we get 
$\sigma_0=7.4039(19)$, $\sigma_0 \tilde a=-0.04(10)$, $\sigma_0 \tilde b=-19.8(9)$, $\tilde d =2.0(1.2)$, and $\chi^2/$d.o.f.$=1.099$.  
For $\sqrt{\sigma} L  \gtrapprox 6$, 
$0.392 < \beta \le 0.4$ we get $\sigma_0= 7.4054(34)$, 
$\sigma_0 \tilde a=-0.12(9)$, $\sigma_0 \tilde b=-19.05(50)$, 
$\tilde d=0.7(3)$, and $\chi^2/$d.o.f.$=1.048$.  We observe that the results
obtained from these two disjoint data sets are consistent. The amplitude 
$\sigma_0 \tilde a$ is consistent with zero at the level of our statistical 
accuracy, as expected for the improved model.

As central value of $\sigma_0$ we have taken the result of the fit with 
$\sqrt{\sigma} L  \gtrapprox 6$ and $\beta \le 0.392$:
\begin{equation}
\sigma_0 = 7.404(5) + 6700 \times (\beta_c - 0.387721735) \;\;.
\end{equation}
The error is taken such that also the results of other fits, in particular 
the one with $0.392 < \beta \le 0.4$, are covered. The dependence on 
the value of $\beta_c$ is estimated by redoing the fit for 
$\beta \le 0.392$ with a slightly shifted value of $\beta_c$.  Note that 
the  dependence on the value of  $\beta_c$ becomes weaker, when data for 
larger values of $\beta$ are fitted.
The dependence on $\nu$ and $\omega$ is small, 
and can be ignored at our level of accuracy.  Combining the estimate of 
$\sigma_0$ 
and that of the amplitude of the second moment correlation length in the 
high temperature phase, eq.~(\ref{xi0}),  we arrive at
\begin{equation}
R_{2nd,+} = \sigma_0 f_{2nd,+}^2  = 0.3863(6) \;.
\end{equation}

Fitting all our data for $\sqrt{\sigma} L  \gtrapprox 6$ and 
$\beta \le 0.4$ we arrive at 
\begin{equation}
\label{sigmacurve} 
 \sigma(t) = 7.40535 (-t)^{1.2599418} \;\left[1  -0.011 \; (-t)^{0.52267}  
 + 1.4352 t  \right] \;,
\end{equation}
where $t=0.387721735 - \beta$.  Below we shall see that this  parametrizes
the interface tension in the interval $0.389 \le \beta \le 0.4$ quite well.
Throughout, the deviation from the true value should be less than or equal to
$3 \times 10^{-7}$ as shown in table \ref{Rresults}. 

\subsection{Analysing the interface free energy}
Finally we computed $R_{2nd,-}$,  $R_{exp,-}$  and the constant $C_0$, 
eq.~(\ref{C0RG}).
To this end, we take the interface tension $\sigma$ computed at the values of
$\beta$ that we had simulated at in ref. \cite{mytwopoint}.  As Ansatz we
used $F_s = \sigma L^2  + c_0 -\frac{1}{4 \sigma L^2}$ with $\sigma$ and 
$c_0$ as free parameters of the fit. We took data obtained for
$\sqrt{\sigma} L  \gtrapprox 6$ into account. Only in the case 
of $\beta=0.389$, our smallest linear size $L=128$ does not satisfy this 
criterion.  We checked possible effects at $\beta=0.39158$, 
where we get a similar value of $\sqrt{\sigma} L$ for $L=64$ as for $L=128$
at $\beta=0.389$.
Comparing the results for the pair of lattice sizes $L=64$
and $128$ with that for $L=128$ and $256$,  we conclude that the 
possible systematical error is smaller than our statistical error 
at $\beta=0.389$.   
Our results are summarized in table \ref{Rresults}.  For $\beta=0.389$, 
$0.39$, and $0.392$ the pair $L=128$ and $256$ of lattice sizes is used,
for $\beta=0.393$ and $0.394$ the lattice sizes $L=64$, $128$ and $256$
enter the fit and for $\beta=0.396$ and $0.4$, the sizes $L=64$ and $128$
are used.

\begin{table}
\caption{\sl \label{Rresults}
In the first column we give the inverse temperature $\beta$. In the second
column our estimates of $\sigma$ obtained from the analysis of the interface
tension. In $()$ we give the statistical error, while the numbers in $[]$ 
give the value of $\sigma$ minus the estimate obtained from 
eq.~(\ref{sigmacurve}). In column 3 we give $c_0 + \frac{1}{2} \ln \sigma$ and
in columns 4 and 5 the products $\sigma \xi_{2nd}^2$ and $\sigma \xi_{exp}^2$,
respectively. The numbers for $\xi_{2nd}$ and $\xi_{exp}$ are taken from 
table II of ref. \cite{mytwopoint}. 
}
\begin{center}
\begin{tabular}{lllll}
\hline
 \mc{1}{c}{$\beta$} & 
\mc{1}{c}{$\sigma$} & 
\mc{1}{c}{ $c_0 + 0.5 \ln \sigma$} & 
\mc{1}{c}{ $\sigma \xi_{2nd}^2$} & 
\mc{1}{c}{ $\sigma \xi_{exp}^2$} \\
\hline
0.4   & 0.02842920(50)[$+29$] &  0.3893(28) & 0.101796(6) &  0.108488(37) \\
0.396 & 0.01740543(34)[$\pm 0$]& 0.3908(19) & 0.102230(6) &  0.108280(36) \\ 
0.394 & 0.01232175(15)[$-22$] &  0.3923(11) & 0.102418(6) &  0.108112(37) \\
0.393 & 0.00991738(13)[$-16$] &  0.3916(10) & 0.102497(6) &  0.108091(36) \\
0.392 & 0.00762273(17)[$-31$] &  0.3955(44) & 0.102570(7) &  0.107921(36) \\
0.391 & 0.00545890(14)[$-9$]  &  0.3900(36) & 0.102664(7) &  0.107885(37) \\
0.39  & 0.00345657(13)[$-11$] &  0.3909(32) & 0.102734(7) &  0.107876(37) \\
0.389 & 0.00167139(11)[$-13$] &  0.3910(25) & 0.102784(11)&  0.107786(54) \\ 
\hline
\end{tabular}
\end{center}
\end{table}

The estimates of $c_0 + 0.5 \ln \sigma$ are constant within the range
of $\beta$-values that we have studied.  As our final result we quote
\begin{equation}
 C_0 = 0.391(2) \;\;,
\end{equation}
which is the average of the estimates for $\beta=0.389, 0.39, 0.391$ and 
$0.392$. Our result is fully consistent with $C_0 = 0.3895(8)$ obtained
in ref. \cite{CaHaPa07} studying the Ising model. The result of 
\cite{CaHaPa07} is more accurate, since interface free energies for 
smaller values of $\sqrt{\sigma} L$ were included in the analysis.

\begin{figure}
\begin{center}
\includegraphics[width=14.5cm]{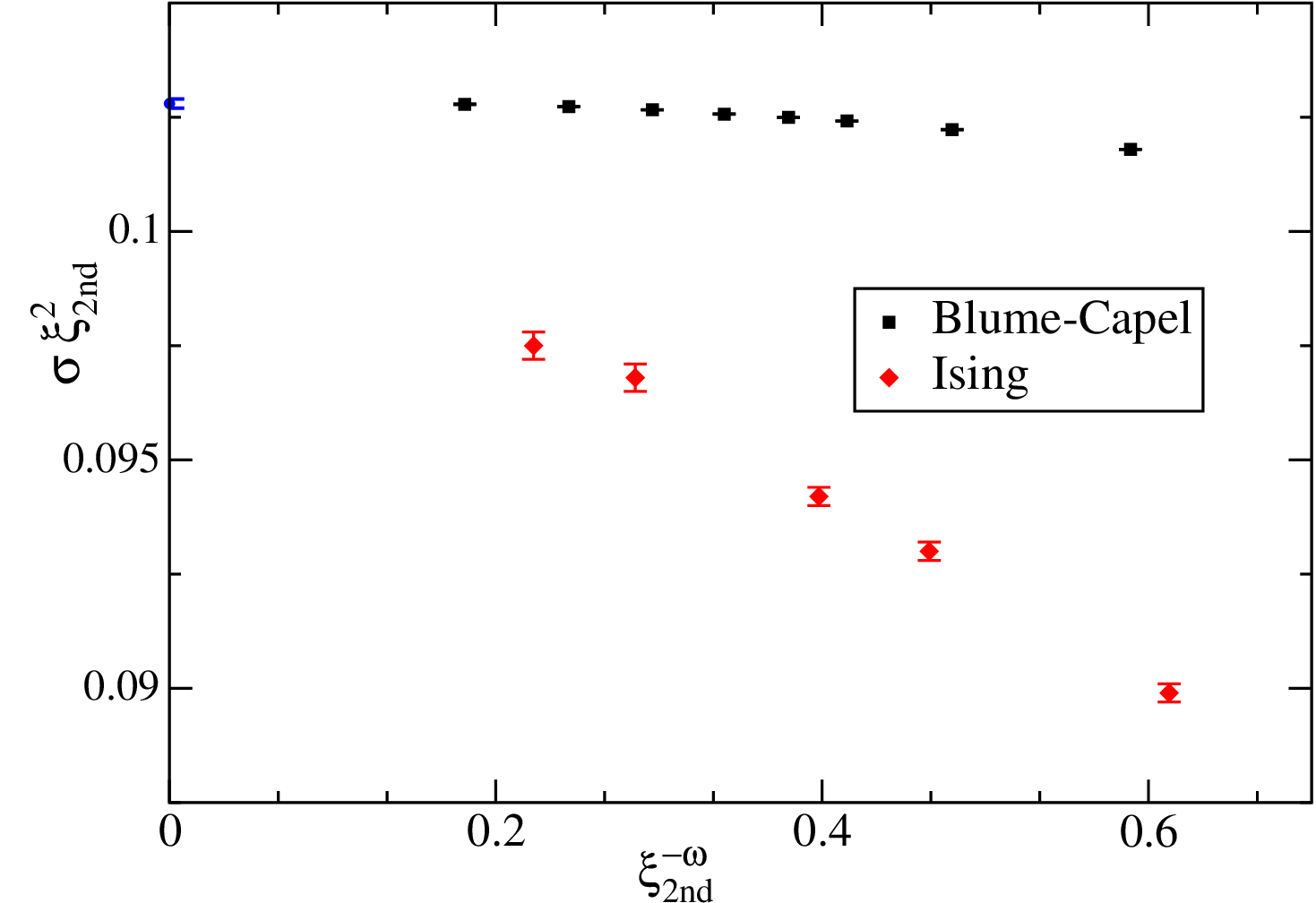}
\caption{\label{R2ndm}
We plot $\sigma \xi_{2nd}^2$  as a function of $\xi_{2nd}^{-\omega}$.
The data for the Ising model are taken from ref. \cite{CaHaPa07}. The filled
circle on the $y$-axis gives our  estimate of $R_{2nd,-}$.
}
\end{center}
\end{figure}

In figure \ref{R2ndm} we plot $\sigma \xi_{2nd}^2$ as a function of 
$\xi_{2nd}^{-\omega}$.
For comparison we plot the corresponding results for the three-dimensional
Ising model given in table 11 of ref. \cite{CaHaPa07}. 

We analysed our numerical results by using the Ansatz
\begin{equation}
 \sigma \xi^2 = R_{-}  + a \xi^{-\omega} + b \xi^{-2}  \;. 
\end{equation}
Fitting with different ranges of the data and also skipping the 
term $a \xi^{-\omega}$ we arrive at the estimates
\begin{equation}
R_{2nd,-} = 0.1028(1) 
\end{equation}
and 
\begin{equation}
R_{exp,-} = 0.1077(3) \;,
\end{equation}
where the error bar is chosen such that the results of various fits are
covered. 

\begin{table}
\caption{\sl \label{Literature}
}
\begin{center}
\begin{tabular}{rrrr}
\hline
 \mc{1}{c}{ref.} &
\mc{1}{c}{$R_{2nd,+}$} &
\mc{1}{c}{$R_{2nd,-}$} &
\mc{1}{c}{$R_{exp,-}$} \\
\hline
\cite{AgCaCaHa97} &            &           & 0.1056(19) \\
\cite{Klaus97}  &            & 0.1040(8) &            \\
\cite{CaHaPa07} & 0.387(2)   & 0.1024(5) & 0.1084(11) \\
\hline
present         & 0.3863(6)  & 0.1028(1) & 0.1077(3) \\
\hline
\end{tabular}
\end{center}
\end{table}

\section{Comparison with results given in the literature}
In experiments on binary mixtures the correlation length can be determined
accurately only in the high temperature phase. Therefore only estimates of 
$R_{2nd,+}$ are available. Reviews of experimental results are given
in  \cite{Moldover85,ChMoSc86}.
In their table I, the authors of \cite{ChMoSc86} summarize results for various
binary liquid mixtures. As mean value they quote $R_+=0.386$ without error bar.
This is also the favoured value of ref. \cite{Moldover85}.
Given the scattering of the data, the error might be a 2 or 3 on the second 
digit. A bit more recently $R_{+}=0.41(4)$ was obtained from the
study of a cyclohexane-aniline mixture in \cite{MaWo96}.
Previous experimental results are summarized in \cite{MaWo96} as 
$R_{+}=0.37(3)$. Our result is nicely consistent with the experimental ones,
confirming that the phase transition of the binary liquid mixtures belongs
to the Ising universality class.

Theoretical estimates of $R_{\pm}$ have been computed by using 
various methods. Br\'ezin and Feng \cite{BrFe84}  computed $R_{2nd,-}$
to order $\epsilon^2$ in the $\epsilon$-expansion. The numerical evaluation 
of their result for $\epsilon=1$ gives results in the range from $\approx 0.051$
up to $\approx 0.057$. Compared with our results, this 
is too small by a factor of about 2.  M\"unster \cite{Muenster} performed a 
semiclassical calculation at one-loop level, which was extended to two-loop
in ref. \cite{Hoppe}.  Their central result is given in eq.~(36) of \cite{Hoppe}
\begin{equation}
R_{2nd,-} = \frac{2}{u^*_R}  \left\{1 + \sigma_{1l} \frac{u^*_R}{4 \pi} 
+ \sigma_{2l} \left( \frac{u^*_R}{4 \pi} \right)^2 + O\left(u_R^{*\; 3} \right) \right\} \;,
\end{equation}
where $\sigma_{1l} = -0.2002602...$ and $\sigma_{2l} = -0.0076(8)$. Plugging 
in $u_R^*=14.08(1)$ \cite{mytwopoint}, we arrive at $R_{2nd,-} = 0.1088(2)$, 
where the number in brackets gives the error  due to the errors of $\sigma_{2l}$ and $u_R^*$.  
The error due to the truncation of the series is hard to estimate.
The result given in table 1 of \cite{Hoppe} for various resummation schemes
might suggest that the error is in the third digit.

In the literature one can find a number of Monte Carlo studies
of the Ising model on the simple cubic lattice.  In table \ref{Literature}
we give results obtained in the last two decades.  Our present results
are consistent with those obtained from simulations of the Ising model, 
but are more accurate.
For a summary of Monte Carlo studies performed before 
1997 see table 8 of ref. \cite{Klaus97}.  

\section{Summary and conclusions}
We have studied the behaviour of the interface free energy in the improved
three-dimensional Blume-Capel model. The interface free energy is determined  
by the difference of the free energy of a system with anti-periodic 
and a system with periodic boundary conditions. For the precise definition see
eqs.~(\ref{fs1},\ref{fs2}). We computed the interface free energy 
by integrating 
the interface energy $E_s$ over the inverse temperature $\beta$ numerically.  
The interface free energy $F_s$ at the starting point of the integration was 
determined by using the boundary flip algorithm \cite{BF}.
The interface energy $E_s$ was computed by using a variance reduced estimator 
based  on the exchange cluster update \cite{ReMaCh98,ChMaRe98}. 
Compared with the standard estimator, the square of the statistical error is
reduced by a factor of up to 70. This finding is in line with refs. 
\cite{mysphere,myfilm}, where we demonstrated that the exchange cluster update
allows to define variance reduced estimators of quantities related to the 
critical Casimir effect. It seems likely that the exchange cluster update
allows to define variance reduced estimators for a wide range of quantities 
related to defects in $\mathbb{Z}_2$-invariant systems.

The dependence of the interface free energy on the transversal extensions
is well described by effective interface models. Recently there had been 
progress in the understanding of the predictive power of these models. 
See for example refs. \cite{Aha,BrMe16} and refs. therein. For 
a more detailed discussion see sec. \ref{effectiveT}.  Here we only used 
these results to extract the interface tension $\sigma$ and the constant $c_0$,
eq.~(\ref{areaExp}), from our data for the interface free energy. In order
to probe the predictions of effective  interface models very accurate data
for a range of interface areas would be needed.

Using our estimates for the interface tension and the results for the 
correlation length obtained in ref. \cite{mytwopoint}, 
we computed the universal amplitude ratios $R_{2nd,+}$, $R_{2nd,-}$, and
$R_{2nd,+}$ with high accuracy.  Our estimate of $R_{2nd,+}$ coincides with 
estimates obtained from experiments on binary liquid mixtures 
\cite{Moldover85,ChMoSc86,MaWo96}. There is also good agreement for all 
three quantities with estimates obtained for the Ising model on the 
simple cubic lattice \cite{AgCaCaHa97,Klaus97,CaHaPa07}. These findings confirm
the universality hypothesis.

\section{Acknowledgement}
This work was supported by the Deutsche Forschungsgemeinschaft under the 
grants No HA 3150/3-1 and HA 3150/4-1.

\end{document}